
%
%
%
%
%
%
\hsize=6.5truein
\vsize=9.0truein

\baselineskip=1.2\baselineskip
\tolerance=10000


\catcode`\@=11 

\def\nolabels{\def\wrlabel##1{}\def\eqlabel##1{}\def\reflabel##1{}}
\def\writelabels{\def\wrlabel##1{\leavevmode\vadjust{\rlap{\smash%
{\line{{\escapechar=` \hfill\rlap{\sevenrm\hskip.03in\string##1}}}}}}}%
\def\eqlabel##1{{\escapechar-1\rlap{\sevenrm\hskip.05in\string##1}}}%
\def\thlabel##1{{\escapechar-1\rlap{\sevenrm\hskip.05in\string##1}}}%
\def\reflabel##1{\noexpand\llap{\noexpand\sevenrm\string\string\string##1}}}
\nolabels
\global\newcount\secno \global\secno=0
\global\newcount\meqno \global\meqno=1
\global\newcount\mthno \global\mthno=1
\global\newcount\mexno \global\mexno=1
\global\newcount\mquno \global\mquno=1
\global\newcount\tblno \global\tblno=1
\def\newsec#1{\global\advance\secno by1 
\global\subsecno=0\xdef\secsym{\the\secno.}\global\meqno=1\global\mthno=1
\global\mexno=1\global\mquno=1\global\figno=1\global\tblno=1

\bigbreak\medskip\noindent{\bf\the\secno. #1}\writetoca{{\secsym} {#1}}
\par\nobreak\medskip\nobreak}
\xdef\secsym{}
\global\newcount\subsecno \global\subsecno=0
\def\subsec#1{\global\advance\subsecno by1 \global\subsubsecno=0
\xdef\subsecsym{\the\subsecno.}
\bigbreak\noindent{\bf\secsym\the\subsecno. #1}\writetoca{\string\quad
{\secsym\the\subsecno.} {#1}}\par\nobreak\medskip\nobreak}
\xdef\subsecsym{}
\global\newcount\subsubsecno \global\subsubsecno=0
\def\subsubsec#1{\global\advance\subsubsecno by1
\bigbreak\noindent{\it\secsym\the\subsecno.\the\subsubsecno.
                                   #1}\writetoca{\string\quad
{\the\secno.\the\subsecno.\the\subsubsecno.} {#1}}\par\nobreak\medskip\nobreak}
\global\newcount\appsubsecno \global\appsubsecno=0
\def\appsubsec#1{\global\advance\appsubsecno by1 \global\subsubsecno=0
\xdef\appsubsecsym{\the\appsubsecno.}
\bigbreak\noindent{\it\secsym\the\appsubsecno. #1}\writetoca{\string\quad
{\secsym\the\appsubsecno.} {#1}}\par\nobreak\medskip\nobreak}
\xdef\appsubsecsym{}
\def\appendix#1#2{\global\meqno=1\global\mthno=1\global\mexno=1
\global\figno=1\global\tblno=1
\global\subsecno=0\global\subsubsecno=0
\global\appsubsecno=0
\xdef\appname{#1}
\xdef\secsym{\hbox{#1.}}
\bigbreak\bigskip\noindent{\bf Appendix #1. #2}
\writetoca{Appendix {#1.} {#2}}\par\nobreak\medskip\nobreak}
%
%
\def\eqnn#1{\xdef #1{(\secsym\the\meqno)}\writedef{#1\leftbracket#1}%
\global\advance\meqno by1\wrlabel#1}
\def\eqna#1{\xdef #1##1{\hbox{$(\secsym\the\meqno##1)$}}
\writedef{#1\numbersign1\leftbracket#1{\numbersign1}}%
\global\advance\meqno by1\wrlabel{#1$\{\}$}}
\def\eqn#1#2{\xdef #1{(\secsym\the\meqno)}\writedef{#1\leftbracket#1}%
\global\advance\meqno by1$$#2\eqno#1\eqlabel#1$$}
%
%
\def\thm#1{\xdef #1{\secsym\the\mthno}\writedef{#1\leftbracket#1}%
\global\advance\mthno by1\wrlabel#1}
\def\exm#1{\xdef #1{\secsym\the\mexno}\writedef{#1\leftbracket#1}%
\global\advance\mexno by1\wrlabel#1}
%
%
\def\tbl#1{\xdef #1{\secsym\the\tblno}\writedef{#1\leftbracket#1}%
\global\advance\tblno by1\wrlabel#1}
%
\newskip\footskip\footskip14pt plus 1pt minus 1pt 
\def\f@@t{\baselineskip\footskip\bgroup\aftergroup\@foot\let\next}
\setbox\strutbox=\hbox{\vrule height9.5pt depth4.5pt width0pt}
\global\newcount\ftno \global\ftno=0
\def\foot{\global\advance\ftno by1\footnote{$^{\the\ftno}$}}
%
\newwrite\ftfile
\def\footend{\def\foot{\global\advance\ftno by1\chardef\wfile=\ftfile
$^{\the\ftno}$\ifnum\ftno=1\immediate\openout\ftfile=foots.tmp\fi%
\immediate\write\ftfile{\noexpand\smallskip%
\noexpand\item{f\the\ftno:\ }\pctsign}\findarg}%
\def\footatend{\vfill\eject\immediate\closeout\ftfile{\parindent=20pt
\centerline{\bf Footnotes}\nobreak\bigskip\input foots.tmp }}}
\def\footatend{}
%
%
\global\newcount\refno \global\refno=1
\newwrite\rfile
\def\ref{\the\refno\nref}
\def\bref{\nref}
\def\nref#1{\xdef#1{\the\refno}\writedef{#1\leftbracket#1}%
\ifnum\refno=1\immediate\openout\rfile=refs.tmp\fi
\global\advance\refno by1\chardef\wfile=\rfile\immediate
\write\rfile{\noexpand\item{[#1]\ }\reflabel{#1\hskip.31in}\pctsign}\findarg}
\def\findarg#1#{\begingroup\obeylines\newlinechar=`\^^M\pass@rg}
{\obeylines\gdef\pass@rg#1{\writ@line\relax #1^^M\hbox{}^^M}%
\gdef\writ@line#1^^M{\expandafter\toks0\expandafter{\striprel@x #1}%
\edef\next{\the\toks0}\ifx\next\em@rk\let\next=\endgroup\else\ifx\next\empty%
\else\immediate\write\wfile{\the\toks0}\fi\let\next=\writ@line\fi\next\relax}}
\def\striprel@x#1{} \def\em@rk{\hbox{}}
\def\lref{\begingroup\obeylines\lr@f}
\def\lr@f#1#2{\gdef#1{\ref#1{#2}}\endgroup\unskip}

\def\addref#1{\immediate\write\rfile{\noexpand\item{}#1}} 
\def\startrefs#1{\immediate\openout\rfile=refs.tmp\refno=#1}
\def\xref{\expandafter\xr@f}\def\xr@f[#1]{#1}
\def\refs#1{[\r@fs #1{\hbox{}}]}
\def\r@fs#1{\edef\next{#1}\ifx\next\em@rk\def\next{}\else
\ifx\next#1\xref #1\else#1\fi\let\next=\r@fs\fi\next}
%

%
 \newwrite\ffile\global\newcount\figno \global\figno=1
%
%
\def\fig{\the\figno\nfig}
\def\nfig#1{\xdef#1{\secsym\the\figno}%
\writedef{#1\leftbracket \noexpand~\the\figno}%
\ifnum\figno=1\immediate\openout\ffile=figs.tmp\fi\chardef\wfile=\ffile%
\immediate\write\ffile{\noexpand\medskip\noexpand\item{Figure\ \the\figno. }
\reflabel{#1\hskip.55in}\pctsign}\global\advance\figno by1\findarg}
\def\xfig{\expandafter\xf@g}\def\xf@g \penalty\@M\ {}
\def\figs#1{figs.~\f@gs #1{\hbox{}}}
\def\f@gs#1{\edef\next{#1}\ifx\next\em@rk\def\next{}\else
\ifx\next#1\xfig #1\else#1\fi\let\next=\f@gs\fi\next}
%
%
\newwrite\lfile

{\escapechar-1\xdef\pctsign{\string\%}\xdef\leftbracket{\string\{}
\xdef\rightbracket{\string\}}\xdef\numbersign{\string\#}}

\def\writestop{\def\writestoppt{\immediate\write\lfile{\string\pageno%
\the\pageno\string\startrefs\leftbracket\the\refno\rightbracket%
\string\def\string\secsym\leftbracket\secsym\rightbracket%
\string\secno\the\secno\string\meqno\the\meqno}\immediate\closeout\lfile}}
\def\writestoppt{}\def\writedef#1{}

\def\seclab#1{\xdef #1{\the\secno}\writedef{#1\leftbracket#1}\wrlabel{#1=#1}}

\def\subseclab#1{\xdef #1{\secsym\the\subsecno}%
\writedef{#1\leftbracket#1}\wrlabel{#1=#1}}
\def\appsubseclab#1{\xdef #1{\secsym\the\appsubsecno}%
\writedef{#1\leftbracket#1}\wrlabel{#1=#1}}
\def\subsubseclab#1{\xdef #1{\secsym\the\subsecno.\the\subsubsecno}%
\writedef{#1\leftbracket#1}\wrlabel{#1=#1}}
\newwrite\tfile \def\writetoca#1{}
\def\leaderfill{\leaders\hbox to 1em{\hss.\hss}\hfill}
\def\writetoc{\immediate\openout\tfile=toc.tmp
   \def\writetoca##1{{\edef\next{\write\tfile{\noindent ##1
   \string\leaderfill {\noexpand\number\pageno} \par}}\next}}}

%
\catcode`\@=12 
%
%
%
%
%
\def\dbend{{\manual\char127}}
\def\d@nger{\medbreak\begingroup\clubpenalty=10000
    \def\par{\endgraf\endgroup\medbreak} \noindent\hang\hangafter=-2
    \hbox to0pt{\hskip-\hangindent\dbend\hfill}\ninepoint}
\outer\def\danger{\d@nger}

\def\darr#1{\raise1.5ex\hbox{$\leftrightarrow$}\mkern-16.5mu #1}
\def\half{{\textstyle{1\over2}}} 

%
%
\def\al{\alpha}
\def\be{\beta}
\def\ga{\gamma}  
\def\de{\delta}  \def\De{\Delta}
\def\ep{\epsilon}

\def\th{\theta}

\def\la{\lambda} \def\La{\Lambda}
\def\rh{\rho}
\def\si{\sigma}

  \def\Ph{\Phi}

\def\om{\omega}  
%
%

%

%
%
 
\def\cC{{\cal C}} 
\def\cE{{\cal E}} \def\cF{{\cal F}}

\def\cU{{\cal U}} 

\def\cW{{\cal W}}

\def\proof{\noindent {\it Proof:}\ }
\def\Box{\hbox{$\rlap{$\sqcup$}\sqcap$}}

\def\Ep{{\cal E}}

%
%
\def\amsyes{y }

\def\answ{y }

\ifx\answ\amsyes
\input amssym.def


\def\CC{{\Bbb C}}
\def\ZZ{{\Bbb Z}}
\def\NN{{\Bbb N}}

\def\bfg{{\frak g}}
\def\bfh{{\frak h}}

   \def\bfbm{{\frak b}_-}
\def\bfnp{{\frak n}_+}
\def\bfnm{{\frak n}_-}
\def\hg{{\widehat{\frak g}}}
\def\whg{\hg}
\def\fC{{\frak C}} 
\def\fH{{\frak H}}

\def\sltw{\frak{sl}_2}  
\def\slth{\frak{sl}_3}

\def\fA{\frak{A}} 
 \def\fP{{\frak P}} 
\def\fH{\frak{H}} 
\def\fC{\frak{C}} 
\else
\def\ZZ{{Z\!\!\!Z}}
\def\CC{{I\!\!\!\!C}}
\def\NN{{I\!\!N}}

\def\bfg{{\bf g}}
\def\bfh{{\bf h}}
\def\bfnm{{\bf n}_-}
\def\bfnp{{\bf n}_+}
\def\hg{\hat{\bf g}}

   \def\bfbm{{\bf b}_-}

\def\sltw{s\ell_2}  
\def\slth{s\ell_3}

\fi
%

%
%
\def\comdiag{
\def\normalbaselines{\baselineskip15pt
\lineskip3pt \lineskiplimit3pt }}
\def\mape#1{\smash{\mathop{\longrightarrow}\limits^{#1}}}

\def\mapne#1{\smash{\mathop{\nearrow}\limits_{\,\,\,\,\,\,#1}}}

\def\mapse#1{\smash{\mathop{\searrow}\limits_{#1\,\,\,\,\,\,}}}

\def\mapcr#1#2{\smash{\mathop{\nearrow\llap{$
\hbox{$\searrow\,\,$}$}}
                  \limits^{\,\,\,\,\,\,#1}_{\,\,\,\,\,\,#2}}}

%
%

\newsymbol\ltimes 226E
\newsymbol\rtimes 226F
%
%
%

\def\AnM#1{Ann.\ Math.\ {\bf #1}}

\def\CMP#1{Comm.\ Math.\ Phys.\ {\bf #1}}

\def\NPB#1{Nucl.\ Phys.\ {\bf B#1}}

\def\PRep#1{Phys.\ Rep.\ {\bf #1}}

%

%
%
\def\SMu{\hbox{\lower 3pt\hbox{ \epsffile{su10.eps}}}}
\def\SMs{\hbox{\lower 3pt\hbox{ \epsffile{ss10.eps}}}}
\def\SMd{\hbox{\lower 3pt\hbox{ \epsffile{sd10.eps}}}}

\def\SMS{\leavevmode\vadjust{\rlap{\smash%
{\line{{\escapechar=` \hfill\rlap{\hskip.3in%
                 \hbox{\lower 2pt\hbox{\epsffile{sd10.eps}}}}}}}}}}
\def\SMH{\leavevmode\vadjust{\rlap{\smash%
{\line{{\escapechar=` \hfill\rlap{\hskip.3in%
                 \hbox{\lower 2pt\hbox{\epsffile{su10.eps}}}}}}}}}}
%
%
\def\LW#1{\lower .5pt \hbox{$\scriptstyle #1$}}
\def\LWr#1{\lower 1.5pt \hbox{$\scriptstyle #1$}}
\def\LWrr#1{\lower 2pt \hbox{$\scriptstyle #1$}}
\def\RSr#1{\raise 1pt \hbox{$\scriptstyle #1$}}

\def\ker{{\rm Ker}}
%


\input tables.tex
\input epsf
\hfuzz=32pt
\nopagenumbers
\pageno=0
\def\mywedge{\textstyle{\bigwedge}}
\def\fA{{\hbox{\rm BV}}}
\def\beo{\be_{\al_1}}
\def\gao{\ga^{\al_1}}
\def\betw{\be_{\al_2}}
\def\gatw{\ga^{\al_2}}
\def\beth{\be_{\al_3}}
\def\gath{\ga^{\al_3}}
\def\bo{b^{-\al_1}}
\def\co{c_{-\al_1}}
\def\btw{b^{-\al_2}}
\def\ctw{c_{-\al_2}}
\def\bth{b^{-\al_3}}

\def\cco{c_1}

\def\cctw{c_2}

\def\tco{\si^{\al_1}}

\def\tctw{\si^{\al_2}}
\def\tbth{\om_{\al_3}}

%
%
%
%
\line{}
\vskip1cm
\centerline{\bf BV-STRUCTURE OF THE COHOMOLOGY OF NILPOTENT SUBALGEBRAS}
\smallskip

\centerline{\bf AND THE GEOMETRY OF ($\cW$-) STRINGS}
\vskip1cm

\centerline{Peter BOUWKNEGT$\,^{1}$, Jim McCARTHY$\,^1$ and
Krzysztof PILCH$\,^2$}
\bigskip

\centerline{\sl $^1$ Department of Physics and Mathematical Physics}
\centerline{\sl University of Adelaide}
\centerline{\sl Adelaide, SA~5005, Australia}
\bigskip

\centerline{\sl $^2$ Department of Physics and Astronomy }
\centerline{\sl  University of Southern California}
\centerline{\sl Los Angeles, CA~90089-0484, USA}
\medskip
\vskip1.5cm

\centerline{\bf ABSTRACT}\medskip
{\rightskip=1cm
\leftskip=1cm
\noindent
Given a simple, simply laced, complex
Lie algebra $\bfg$ corresponding to the
Lie group $G$, let $\bfnp$ be the subalgebra generated by the positive
roots.  In this paper we construct a
BV-algebra $\fA[\bfg]$ whose underlying graded
commutative algebra is given by the cohomology, with respect to $\bfnp$,
of the algebra of regular functions on $G$ with values in
$\mywedge (\bfnp\backslash\bfg)$. We conjecture that $\fA[\bfg]$
describes the algebra of {\it all} physical (i.e., BRST invariant)
operators of the noncritical $\cW[\bfg]$ string.  The
conjecture is verified in the two explicitly
known cases, $\bfg=\sltw$ (the Virasoro string) and $\bfg=\slth$ (the
$\cW_3$ string).
}

\vfil
\line{USC-95/32 \hfil}
\line{ADP-95-59/M41 \hfil}
\line{{{\tt hep-th/9512032}}\hfil December 1995}

\eject

\bref\Wi{
E.~Witten,\NPB{373} (1992) 187 {\tt (hep-th/9108004)}.}

\bref\WiZw{
E.~Witten and B.~Zwiebach, \NPB{377} (1992) 55
{\tt (hep-th/9201056)}.}

\bref\LZbv{
B.H.~Lian and G.J.~Zuckerman,  \CMP{154} (1993) 613
{\tt (hep-th/9211072)}.}

\bref\MS{
M.~Penkava and A.~Schwarz, in ``Perspectives in Mathematical
Physics,'' Vol.\ III, {\it eds.} R.~Penner and S.T.~Yau,
(International Press, 1994) {\tt (hep-th/9212072)}.}

\bref\Get{
E.~Getzler, \CMP{159} (1994) 265 {\tt (hep-th/9212043)}. }

\bref\BMPbig{
P.~Bouwknegt, J.~McCarthy and K.~Pilch, {\it The $\cW_3$ algebra:
modules, semi-infinite cohomology and BV-algebras}, preprint
USC-95/18, ADP-46/M38 ({\tt hep-th/9509119}).}

\bref\BGG{
I.N.~Bernstein, I.M.~Gel'fand and S.I.~Gel'fand,  in
``Lie groups and their representations,'' Proc.\ Summer School in
Group Representations, Bolyai Janos Math.\ Soc., Budapest 1971, pp.\
21, (Halsted, New York, 1975).}

\bref\LZvir{
B.H. Lian and G.J. Zuckerman, Phys.\ Lett.\ {\bf 266B} (1991) 21.}

\bref\BMPvir{
P.~Bouwknegt, J.~McCarthy and K.~Pilch,  \CMP{145} (1992)
541.}

\bref\BK{
R.~Bott, \AnM{66} (1957) 203;  B.~Kostant, \AnM{74} (1961) 329. }

\bref\BMPc{
P.~Bouwknegt, J.~McCarthy and K.~Pilch,  in the proceedings of the workshop
``Quantum Field Theory and String Theory,'' Carg\`ese 1993,
{\it eds.\ } L.~Baulieu et al., p.\ 59
(Plenum Press, New York, 1995) {\tt (hep-th/9311137)}.}

\bref\FF{
E.~Feigin and E.~Frenkel, \CMP{128} (1990) 161.}

\bref\Kac{
V.G.~Kac, {\it Infinite dimensional Lie algebras}, (Cambridge
University Press, Cambridge, 1990).}

\bref\BMPnp{
P.~Bouwknegt, J.~McCarthy and K.~Pilch, \NPB{377} (1992) 541
{\tt (hep-th/9112036)}.}

\bref\BMPcmp{
P.~Bouwknegt, J.~McCarthy and K.~Pilch, \CMP{131} (1990) 125.}

\bref\BS{
P.~Bouwknegt and K.~Schoutens, \PRep{223}
(1993) 183 {\tt (hep-th/9210010)}.}

\footline{\hss \tenrm -- \folio\ -- \hss}

\newsec{Introduction}
\seclab\SSintro

 In the past few years there has been a considerable effort to
understand algebraic and geometric structures underlying $\cW$-gravity
in two dimensions.
The initial observation [\Wi,\WiZw] was that a subsector of the
algebra of physical (i.e., BRST invariant) operators $\fH[\cW_2]$ of
the $\cW_2$ (Virasoro) string is modeled by polynomial polyvectors on
the complex plane.  Subsequent work [\LZbv,\MS,\Get] revealed that
the full algebra of physical operators has the structure of a
Batalin-Vilkovisky (BV-) algebra, with the BV-algebra of polyvectors
arising as a quotient determined by the natural action of $\fH[\cW_2]$
on the ground ring.\medskip

 In a recent paper [\BMPbig] we found a similar, albeit
significantly more complicated, description of the operator algebra of
the $\cW_3$ string. Crucial to our construction was the observation
that the proper geometric framework for studying the $\cW_n$ string
($n=2,3$) is the base affine space [\BGG] of $SL(n)$. The
operator algebra, $\fH[\cW_n]$, can then be parametrized as a direct
sum of (twisted) polyderivations of the ground ring, and contains the
BV-algebra of polyvectors on the base affine space as its quotient.
However, a simple characterization of the full BV-algebra of the
$\cW_n$ string in terms of geometric objects associated with
$\frak{sl}_n$ and its base affine space was not known.\medskip

 In this letter we will provide the desired characterization.  We will
show that with any (simply-laced) Lie algebra $\bfg$ one may associate
a BV-algebra, $\fA[\bfg]$, whose underlying graded commutative algebra
is given by the cohomology, $ H(\bfnp,\cE(G)\otimes \mywedge
(\bfnp\backslash\bfg))$, with respect to the maximal nilpotent
subalgebra $\bfnp\subset \bfg$, where $\cE(G)$ is the space of regular
functions on the complex Lie group $G$ of $\bfg$.  Since the
polyvectors on the base affine space of $G$ are characterized as the
$\bfnp$-invariant elements in
$\cE(G)\otimes\mywedge(\bfnp\backslash\bfg)$, they are incorporated into
this picture as the zero-th order cohomology.  For the $\cW$-strings
whose spectra are explicitly known, the $\cW_2$ string
[\LZvir,\BMPvir] and the $\cW_3$ string [\BMPbig], we verify that the
higher order cohomologies account correctly for the remaining sectors
of the spectrum. This gives a simple geometric picture for the
operator algebra of rather complicated models without recourse to
generalized polyvectors. \medskip

This result is also of interest since it relates the semi-infinite
cohomology of an infinite-dimensional $\cW$-algebra with coefficients
in infinite-dimensional modules (Fock modules), to the Lie algebra
cohomology of a finite-dimensional Lie algebra $\bfnp$ with
coefficients in $L(\La)\otimes
\mywedge(\bfnp\backslash\bfg)$, where $L(\La)$, $\La\in P_+$,  is a
finite-dimensional irreducible module of $\bfg$. (The irreducible
modules arise upon decomposition of $\cE(G)$ with respect to $\bfg$.) To
determine the latter cohomology we are led to a classical problem in
Lie algebra cohomology; namely, the computation of
$H(\bfnp,L(\La))$ [\BK]. Here we will argue that for a
weight $\La\in P_+$ sufficiently deep inside the fundamental Weyl
chamber,
\eqn\KPaa{
H(\bfnp,L(\La)\otimes\mywedge(\bfnp\backslash\bfg)) ~\cong~
H(\bfnp,L(\La)) \otimes\mywedge(\bfnp\backslash\bfg)\,.}
\medskip

The result in this paper is consistent with the important role played
by $\bfnp$-cohomology in $\cW$-gravity noticed earlier [\BMPc] for the
so-called ``generic'' regime of central charge.  The explanation of
these observations for $\cW$-gravities is expected to be found in the
context of Hamiltonian reduction, or $G/G$ models, but their precise
origin is still unknown.\medskip

Throughout this letter we use the conventions and notation
of [\BMPbig].


\newsec{BV-structure of $H(\bfnp,\Ep(G)\otimes \mywedge \bfbm)$}

Let $\bfg$ be a complex, simple,
simply-laced, finite dimensional Lie algebra.  We fix
a Cartan decomposition $\bfg \cong \bfnm \oplus \bfh \oplus \bfnp
\cong \bfbm \oplus \bfnp$ and denote  the corresponding
Chevalley generators of $\bfg$ by $\{ e_{-\al}, h_i, e_\al\}$,
$i=1,\ldots,\ell \equiv {\rm rank\ }\bfg$, $\al\in\De_+$. The collective
indices of $\bfbm$ and $\bfg$ will be denoted by $a=(-\al,i)$ and
$A=(-\al,i,\al)$, respectively, and the structure constants of $\bfg$
in the chosen basis by $f_{AB}{}^C$. We use the summation convention
in which the repeated indices are summed over their ranges. \medskip

The space, $\cE(G)$, of regular functions on the complex Lie group $G$
of $\bfg$ carries the left and right regular representations of
$\bfg$ corresponding to the left and right invariant vector fields on
$G$. Denote the operators representing the action of a generator $e_A$
by $\Pi^R_A$ and $\Pi^L_A$, respectively. The Peter-Weyl theorem
asserts that as a $\bfg_L\oplus\bfg_R$ module
\eqn\KPpewl{
\cE(G)\cong\bigoplus_{\La\in P_+} L(\La^*)\otimes L(\La)\,,}
where $L(\La)$ and $L(\La^*)$ are the  irreducible
finite dimensional $\bfg$-modules with highest (dominant integral)
weights $\La$ and $\La^*$, respectively, and $\La^*=-w_0\La$. \medskip

Following [\BGG], we define the base affine space of $G$ as the
quotient $A=N_+\backslash G$, where $N_+$ is the subgroup generated by
$\bfnp$. The aim of this section is to study geometric objects that
are associated with $A$.  The basic example is the space of
polyvectors, $\fP(A)$ defined as regular sections of the homogenous
vector bundle $G\times_{N_+} \bigwedge \bfbm$.  Here we consider
$\mywedge \bfbm$ as an $\bfnp$-module through the identification
$\mywedge \bfbm \cong \mywedge ( \bfnp \backslash\bfg)$.  Clearly there
is a natural grading $\fP(A)=\bigoplus_n\fP^n(A)$ induced from the
decomposition
$$
\mywedge \bfbm ~\cong~ \bigoplus_{n=0}^{D}\ \mywedge^n \bfbm\,,\quad\quad
D~=~|\De_+| +\ell\,.
$$

At this point it convenient to introduce a set of ghost oscillators
$\{ b^a, c_a\}$, corresponding to $\bfbm$, with nontrivial
anti-commutators $[b^a , c_b ] = \de^a{}_{b}$ and associated ghost
Fock space $F^{bc}$ with vacuum $|bc\rangle$ satisfying $b^a
|bc\rangle =0$.  We identify $\mywedge \bfbm$ with $F^{bc}$, with the
$\bfnp$ action given by
$\Pi^{bc}_\al = f_{\al b}{}^{c}\, c_c\, b^b$.
In particular this identification induces a graded commutative
product on $F^{bc}$.  Moreover, $F^{bc}$ is also an
$\bfh$-module with the $\bfh$ generators
$\Pi^{bc}_i = f_{i b}{}^c \, c_c \,b^b$.\medskip

Let  $\cE(G) \otimes \bigwedge \bfbm$
denote the space of regular functions on $G$ with values in
$\bigwedge\bfbm$. It has a natural structure of a graded, graded
commutative algebra and carries  commuting actions of $\bfg\oplus\bfh$
and $\bfnp$, defined by the operators $\Pi_A^R$ and $\Pi_i^L+\Pi^{bc}_i$,
and $\Pi_\al^L+\Pi^{bc}_\al$, respectively. Both $\bfg\oplus \bfh$ and
$\bfnp$ act by derivations of the algebra product. In terms of
$\cE(G)\otimes\bigwedge \bfbm$ the polyvectors $\fP(A)$ are simply given by
the $\bfnp$-invariant elements, i.e.,
\eqn\KPpol{
\fP^n(A)~\cong~(\cE(G)\otimes \mywedge^n\bfbm)^{\bfnp}\,.
}
We note that polyvectors of order $0$ are simply identified with the
regular functions, $\cE(A)$, on $A$. Using \KPpewl, the decomposition
of $\cE(A)$ with respect to $\bfh\oplus \bfg$ yields
\eqn\KPmod{
\cE(A) ~\cong~ \bigoplus_{\La\in P_+} \CC_{\La^*}\otimes
L(\La)\,,
}
i.e., $\cE(A)$ is a model space of $\bfg$.
The computation of higher order polyvectors is more involved due to
the typically reducible, but indecomposable, action of $\bfnp$ on
$\bigwedge\bfbm$ (see  [\BGG,\BMPbig]). \medskip

The natural framework for determining invariants of group actions is
Lie algebra cohomology. In this more general context, the
polyvectors are  obtained as the zero-th order cohomology of $\bfnp$ with
coefficients in $\cE(G) \otimes \bigwedge \bfbm$.  One is naturally led to
consider the algebra, $\fA[\bfg]$, defined by the full cohomology,
\eqn\JMdec{
\fA[\bfg] ~\equiv~ H(\bfnp,\cE(G)\otimes  \mywedge \bfbm)\,.
}

 Let us now examine $\fA[\bfg]$ more closely, using this as an opportunity
to introduce further notation and to derive some elementary results.
We introduce a set of ghost oscillators $\{ \si^\al, \om_\al\}$,
corresponding to $\bfnp$, with nontrivial anti-commutators
$[\si^\al , \om_\be ] = \de^\al_{\be}$ and associated ghost Fock space
$F^{\si\om}$ with vacuum $|\si\om\rangle$ satisfying
$\om_\al |\si\om\rangle = 0$.  The $\bfnp$ action on $F^{\si\om}$
is given by
$\Pi^{\si\om}_\al =  -f_{\al\be}{}^\ga\, \si^\be\,\om_\ga$,
and the $\bfh$ action by $\Pi^{\si\om}_i = - f_{i\al}{}^\al\, \si^\al\,
\om_\al$.
Again, there is a natural graded commutative product on $F^{\si\om}$.\medskip

The cohomology $H(\bfnp,\cE(G) \otimes \mywedge \bfbm)$ may now be computed
as the cohomology of the differential
\eqn\eqNPa{
d ~=~ \si^\al
  \left( \Pi^L_\al + \Pi^{bc}_\al + \half \Pi^{\si\om}_\al \right)
}
acting on the complex $\cC(G)
\equiv \cE(G) \otimes F^{bc} \otimes F^{\si\om}$,
i.e.\ $H_{d}(\cC(G))$.  The complex is bi-graded by the $bc$-ghost and
the $\si\om$-ghost numbers (${\rm gh}(c_a) = - {\rm gh}(b^a) = (0,1)$,
$ {\rm gh}(\si^\al) = -{\rm gh}(\om_\al) =(1,0)$), with $d$ of degree
$(1,0)$. Clearly, this bi-degree passes to the cohomology. We will write
$H^n(\bfnp,\cE(G) \otimes \bigwedge\bfbm)$ for
the cohomology in {\it total} ghost number $n$.\medskip

 Combining the above discussions, the complex $\cC(G)$ is a
$\bfg\oplus\bfh$-module under $\Pi^R_A$ and
\eqn\eqNPf{
\Pi_i ~\equiv~ \Pi^L_i + \Pi^{bc}_i + \Pi^{\si\om}_i \,.
} Note that with respect to this $\bfh$-action, the weights of
$b^{-\al}$ and $\om_\al$ are $\al$, whilst those of $\si^\al$ and
$c_{-\al}$ are $-\al$.  Since $d$ commutes both with the action of
$\bfg$ and $\bfh$, we have a direct sum decomposition
\eqn\KPdec{
\fA[\bfg]~=~ \bigoplus_{\La\in P_+}  H(\bfnp,L(\La^*)\otimes \mywedge
\bfbm) \otimes L(\La)\,.
}
As an $\bfh$-module,
\eqn\eqNPh{
H(\bfnp,L(\La) \otimes \mywedge \bfbm) ~\cong~ \bigoplus_{\la \in
P(\cC(\La))} \ H(\bfnp,L(\La) \otimes \mywedge \bfbm)_\la\,, } where,
obviously,  $\cC(\La)=L(\La) \otimes F^{bc} \otimes
F^{\si\om}$, and $P(V)$ denotes the set of
weights of an $\bfh$-module V.\medskip

 The decomposition \KPdec\ reduces the problem of computing $\fA[\bfg]$
to that of computing cohomology of finite-dimensional modules. We
postpone a more detailed discussion till Section 3 and first concentrate
on global properties of $\fA[\bfg]$.

\thm\KPgral
\proclaim Lemma \KPgral. $\fA[\bfg]$ is a graded, graded commutative
algebra with the product ``$\,\cdot\,$'' induced from the  product
on the underlying complex $\cC(G)$.
\smallskip

\proof Let $|0\rangle=|bc\rangle\otimes |\si\om\rangle$. Then $\Ph\in
\cC(G)$ is of the form
\eqn\KPelem{
\Ph=\Ph^{a_1\ldots a_{m}}{}_{\al_1\ldots \al_{n}}
\si^{\al_1}\ldots \si^{\al_{n}} c_{a_1}\ldots
c_{a_{m}}|0\rangle\,,\qquad
\Ph^{a_1\ldots a_{m}}{}_{\al_1\ldots \al_{n}}\in\cE(G)\,.}
The product of two such element is thus given by
\eqn\KPprod{
\Ph
\cdot
\Psi ~=~
(-1)^{np}\,
\Ph^{a_1\ldots a_{m}}{}_{\al_1\ldots \al_{n}}
\Psi^{b_1\ldots b_{p}}{}_{\be_1\ldots \be_{q}}
\si^{\al_1}\ldots\si^{\al_n}\si^{\be_1} \ldots\si^{\be_q} c_{a_1}
\dots c_{a_m}c_{b_{1}}\ldots c_{b_{p}}|0\rangle\,.
}
We verify that
\eqn\KPdonpr{
d(\Ph\cdot\Psi)~=~d\Ph\cdot\Psi+(-1)^{m+n} \Ph\cdot d\Psi\,,
}
from which it follows immediately that the product passes to the
cohomology. Obviously, it is graded commutative according
to the total ghost number.  \Box

The algebra $\fA[\bfg]$ contains the algebra of polyvectors $\fP(A)$
as a subalgebra, namely
\eqn\KPplo{
\fP^m(A)~\cong~\fA^{(0,m)}[\bfg]\,.}
Let $\imath:\fP(A)\longrightarrow \fA[\bfg]$ be the corresponding
embedding. Then we have $\imath(\fP(A))=\bigcap_{\al}\ker\,\om_\al$, and
compatibility with the cohomology requires that $\imath(\fP(A))$ be
$\bfnp$-invariant.\medskip

The space $\fP(A)$ also carries an additional BV-algebra structure,
see e.g.\  [\BMPbig].
We will now prove that there are natural
extensions of the BV-structure from $\fP(A)$ to $\fA[\bfg]$. \medskip

 Recall that a BV-algebra $(\fA,\,\cdot\,,\De)$ consists of a graded,
graded commutative algebra $(\fA,\,\cdot\,)$ with an
operator $\De$, called the BV-operator, which is a graded second order
derivation of degree $-1$ on $\fA$ satisfying  $\De^2=0$. We
refer the reader to [\LZbv,\MS,\Get,\BMPbig] for further details.\medskip

We will introduce the BV-algebra structure on $\fA[\bfg]$ in two
steps. First, we will show that there is a natural extension of the
BV-operator from $\fP(A)$ to $\fA[\bfg]$ that preserves the space of
polyvectors, but has nontrivial cohomology. Secondly, we  will
construct a deformation of the ``naive'' BV-operator such that the
cohomology becomes trivial. It is the latter BV-structure that turns
out to be relevant for $\cW[\bfg]$ strings, as will be discussed in
Section 4.\medskip

\thm\KPfbv
\proclaim Theorem \KPfbv. {Consider the operator
\eqn\KPnaive{
\De_0 ~=~ -b^a(\Pi^L_a+\Pi^{\si\om}_a)+  \half f_{ab}{}^c
b^ab^bc_c\,,}
where $\Pi^{\si\om}_a=-f_{a\al}^\be b^a\si^\al \om_\be$.
Then
\smallskip
\item{i.} $[d,\De_0]=0$,
\item{ii.} $\De_0$ is a BV-operator on $\fA[\bfg]$,
\item{iii.} $\De_0 \imath(\fP(A))\subset\imath(\fP(A))$. }
\smallskip

\proof Note that $-\De_0$ is the differential of
$\bfbm$-cohomology with coefficients in  $\cE(G) \otimes F^{\si\om}$. Thus
$\De_0^2=0$, a fact easily verified by an explicit algebra. Moreover,
if we combine the $\bfbm$ and $\bfnp$ ghosts as $c^A=\{c_a,\om_\al\}$
and $b^A=\{b^a,\si^\al\}$ then
\eqn\KPde{
\de ~=~ d-\De_0=b^A\Pi^L_A-\half f_{AB}^C b^A b^Bc_C\,,
}
is the differential of a twisted cohomology [\FF] of $\bfg$ with
coefficients in $\cE(G)$. Thus $\de^2=0$ and the assertion (i)
follows.\medskip

The second order derivation property of $\De_0$ is shown as follows:
The first term in $\De_0$ is a product of first order derivations
$\Pi^L_a+\Pi^{\si\om}_a$ on $\cE(G)\otimes F^{\si\om}$ and $b^a$ on
$F^{bc}$. The product of such first order derivations acting on the
tensor product of spaces is well-known to be a second order
derivation. The second term, with the nontrivial action on $F^{bc}$
only, upon normal ordering of the $bc$-ghosts becomes a sum of terms
with one or a product of two $b^a$'s, which are first and  second
order derivations, respectively. Since we have already shown that
$\De_0^2=0$, this proves (ii). \medskip

Part (iii) follows from the observation that on
$\imath(\fP(A))$ the only term involving the $\si\om$-ghosts
vanishes. Then $\De_0$ reduces to the BV-operator on $\fP(A)$
constructed in [\BMPbig]. \Box \medskip

We will now seek a deformation of $\De_0$ of the form
$\De=\De_0+\De_1$, such that $\De$ is a BV-operator on $\fA[\bfg]$. In
particular this requires that $[d,\De_1]=0$, which implies that
$\De_1$ must be a nontrivial element in the ``operator cohomology'' of
$d$ on $\cU(\bfg)\otimes \cU(bc)\otimes \cU(\si\om)$. Here
$\cU(\cdot)$ denotes the enveloping algebra and $d$ acts by the
commutator.  Since computing this cohomology is difficult, we make
the further simplification that $\De_1$ is an element of
$\cU(bc)\otimes \cU(\si\om)$. This assumption is motivated by the
naive expectation  that the only second order derivation that has
degree $-1$ and acts nontrivially on the $\cE(G)$ component in
$\fA[\bfg]$ must be of the form $b^a\Pi^L_a$, and thus is already
accounted for in $\De_0$. Then we have

\thm\KPbigbv
\proclaim Theorem \KPbigbv. Let $\bfg$ be a simple, simply laced Lie
algebra and $\ep:Q\times Q\longrightarrow \{\pm1\}$ its asymmetry
function with respect to the chosen Chevalley basis.\foot{{\rm
Recall that $\ep(\al,\be)=f_{\al\be}{}^\ga$, $\al,\be,\ga\in \De$,
see e.g.\ [\Kac].}}
Define
\eqn\KPbvex{
\De'~=~\sum_{\al\in\De_+}\sum_{i=1}^\ell (\al,
\al_i)\,\si^\al b^{-\al} b^i- \sum_{\al,\be\in\De_+} \ep(\al,\be)\si^{\al+\be}
b^{-\al}b^{-\be} \,.
}
Then $\De_t=\De_0+t\De'$ is a one parameter family of BV-operators on
$\fA[\bfg]$.
\smallskip

\proof The required properties are proved by explicit algebra.
The details will be presented in the revised version of
[\BMPbig]. \Box

\medskip
\noindent
{\it Remarks:}
\item{1.} We have verified that for $\bfg=\sltw$ and $\slth$ the generator
$\De'$ of the one parameter family of deformations of $\De_0$ is
uniquely determined by requiring that it be a nontrivial element of
the $\bfnp$-cohomology at ghost number $-1$. It is an interesting open
problem to determine whether this also true for an arbitrary $\bfg$.
\item{2.} The assumption that $\bfg$ is simply laced is merely
technical, and it should be straightforward to obtain a generalization
of \KPbvex\ to the non-simply laced case.
\medskip

A simple scaling argument shows that in fact all BV-algebra structures
on $\fA[\bfg]$ for $t\not=0$ are equivalent. Indeed, if we let
$\si^\al\rightarrow \la\si^\al$, $\om_\al\rightarrow \la^{-1}
\om_\al$ then  $d\rightarrow \la d$,
$\De_0\rightarrow \De_0$ while  $\De'\rightarrow \la \De'$. Since
the cohomology classes must scale homogenously with respect to this
transformation, we may use it to set $t=1$, and denote $\De=\De_{t=1}$.
\medskip

Finally, we have shown in [\BMPbig] that the cohomology of $\De_0$ on
$\fP(A)$ is nontrivial and spanned by the volume element $\prod_a
c_a|bc\rangle$. We now have

\thm\KPconj
\proclaim Conjecture \KPconj.  $\fA[\bfg]$ is acyclic with respect to
$\De$.

\smallskip
This conjecture holds in the case of $\sltw$ and $\slth$, where
$\fA[\bfg]$ can be computed explicitly as we now show.


\newsec{Computation of $H(\bfnp,L(\La) \otimes \mywedge \bfbm)$}

In this section we will compute the cohomology
$H(\bfnp,L(\La) \otimes \mywedge \bfbm)$ for $\La\in P_+$ which,
by \KPdec, implies the result for $H(\bfnp,
\Ep(G) \otimes \mywedge \bfbm)$. \medskip

If $\La$ is sufficiently deep inside the fundamental Weyl chamber
(henceforth we refer to such $\La$ as ``in the bulk'') the cohomology is
easily computed.  The result is given by the following

\thm\thNPa
\proclaim Theorem \thNPa.  Let $\La\in P_+$
\item{i.} The cohomology
$H^n(\bfnp,L(\La) \otimes \mywedge \bfbm)_{\La'}$ is nontrivial
only if there exists a $w\in W$ and $\la\in P(\mywedge^k \bfbm)$ such
that $\La' = w(\La+\rh)-\rh + \la$ and $n=\ell(w) + k$.
\item{ii.} For $\La\in P_+$
in the bulk, i.e.\ $(\La,\al_i)\geq N(\bfg)$ for
some $N(\bfg)\in\NN$ sufficiently large
(in particular $N(\frak{sl}_n) = n-1$), we have
\eqn\eqNPd{
H(\bfnp,L(\La) \otimes \mywedge \bfbm) ~\cong~
  H(\bfnp,L(\La)) \otimes \mywedge \bfbm  ~\cong~
  \left( \bigoplus_{w\in W} \CC_{\,w(\La+\rh)-\rh}
  \right) \otimes \mywedge \bfbm \,.
}

\proof
Consider the gradation of the complex $\cC
= \oplus_{k\in\ZZ}\, \cC_k$ given by $(\rh,\la^{bc})$,
where $\la^{bc}$ is the weight corresponding to $\Pi^{bc}_i$.
With respect to this gradation the differential \eqNPa\
decomposes as
$$
d ~=~ \sum_{k\geq0} \ d_k \,,
$$
with
$$
d_0 ~=~ \si^\al \left(\Pi^L_\al +  \half \Pi^{\si\om}_\al \right)\,,\qquad\quad
d_k ~=~ \sum_{ (\rh,\al) = k } \ \si^\al \Pi^{bc}_\al\,,\qquad k\geq1\,.
$$
In particular, $d_k\equiv0$ for $k\geq (\rh,\th)+1 = {\rm h}^\vee$, where
${\rm h}^\vee$ is the dual Coxeter number of $\bfg$ (for $\bfg =
\frak{sl}_n$ we have ${\rm h}^\vee=n$).
The spectral sequence $(E_k,\de_k)$
corresponding to this gradation converges since
the complex is finite dimensional
(see e.g.\ [\BMPnp] for an elementary exposition).
The first term in the spectral sequence is given by
$$
E_1 ~=~ H_{d_0}(\cC) ~\cong~  H(\bfnp,L(\La)) \otimes \mywedge \bfbm \, ,
$$
where $H(\bfnp,L(\La)) \cong \oplus_{w\in W}\, \CC_{\,w(\La+\rh)-\rh}$
[\BK].  This proves the first part of the theorem.
To prove the second part
we will now determine the condition for the spectral sequence to collapse
at the first term.

The differential $\de_1$ on $E_1$ is simply given by $d_1$, so we find
that a sufficient condition for $\de_1$ to act
trivially is that there exist
no $w,w'\in W\,, \ell(w') = \ell(w) + 1$, $\la,\la' \in P(\mywedge \bfbm)$
such that
$$
w(\La+\rh) - w'(\La+\rh) ~=~ \la' - \la ~=~ \al_i\,,
$$
for some $i$.  Similarly, we find that a sufficient
condition for $\de_k$ to act trivially on $E_k$ is that there exist
no $w,w'\in W\,, \ell(w') = \ell(w) + 1$, $\la,\la' \in P(\mywedge \bfbm)$
such that
$$
w*\La - w'*\La ~=~ \la' - \la ~\in~ \{ \be \in Q_+ \,|\, (\rh,\be)\leq k\}\,.
$$
So, since $\de_k\equiv 0$ for $k\geq {\rm h}^\vee$, we find that a
sufficient condition for $E_\infty \cong \ldots \cong E_2 \cong E_1$ is
that there exist
no $w,w'\in W\,, \ell(w') = \ell(w) + 1$, $\la,\la' \in P(\mywedge \bfbm)$
such that
$$
w*\La - w'*\La ~=~ \la' - \la ~\in~ \{ \be \in Q_+ \,|\, (\rh,\be)\leq
 {\rm h}^\vee -1 \}\,.
$$
This condition is met when $\La$ is sufficiently deep inside the
fundamental Weyl chamber,
i.e.\ $(\La,\al_i) \geq N(\bfg)$ for some $N(\bfg) \in \NN$ sufficiently
large.
\Box\medskip

In principle the cohomology for $\La$ away from the bulk can be
computed by explicitly going through the spectral sequence discussed
above.  For $\sltw$ this is particularly easy since only the
singlet weight $\La=0$ is not in the bulk.  In this case one easily
verifies that the states $c_{-\al}|bc\rangle \otimes |\si\om\rangle$
and $c_1|bc\rangle \otimes \si^\al|\si\om\rangle$ in $E_1$ are eliminated
in $E_2$.  In other words, for $\sltw$ one finds that
$H(\bfnp,L(\La) \otimes \mywedge \bfbm)$ for $\La=0$ contains six states
(organized in three doublets at ghost numbers $0,1$ and $2$ and
$\bfh$-weights $0,-\al,-2\al$, respectively, where a doublet at ghost
number $n$ is a pair of states of the same weight at ghost numbers
$n$ and $n+1$), as opposed to eight states (Theorem \thNPa) for $\La\neq0$.
\medskip

For algebras other than $\sltw$, going through
the above spectral sequence becomes rather cumbersome.  Instead we will
present an independent calculation, based on free field techniques,
for the other case of special interest -- namely,
$\slth$.  Here we will determine the cohomology $H(\bfnp, L(\La)\otimes
\mywedge\bfbm)$,  for arbitrary $\La\in P_+$,
through a spectral sequence associated
with the resolution $(\cF,\delta)$ of the irreducible module
$L(\La)$ in terms of free field
Fock spaces which are co-free over $\bfnp$, i.e.\ isomorphic to
contragradient Verma modules.
The reader should consult [\FF,\BMPcmp] for a review of such
techniques, but for completeness we will recall the little of this theory
which is required here.

Introduce the free oscillators $\be_\al,\ga^\al, \, \al \in \De_+$,
with nontrivial commutators  $[\ga^\al,\be_\be] = \delta^\al{}_\be$ and
associated Fock space $F^{\be\ga}_\La$ with vacuum  $|\La\rangle, \, \La
\in P$, satisfying $\be_\al |\La\rangle = 0$.  Then,
$F^{\be\ga}_\La$ can be given the structure of a
$\bfg$-module in a natural way, with highest weight
$\La$.  For the example of $\slth$,
the positive root generators are realized as
\eqn\real{   \eqalign{
e_{\al_1} & ~=~ \beo \cr
e_{\al_2} & ~=~ \betw -\gao\beth \cr
e_{\al_3} & ~=~ \beth \, , \cr}
}
For $\La \in P_+$ there exists a complex of such Fock modules
$$
0~\longrightarrow~ \cF_{\La}^{(0)}~\buildrel \de^{(0)}\over\longrightarrow~
  \cF_{\La}^{(1)}~\buildrel \de^{(1)}\over\longrightarrow~ \ldots
  ~\buildrel \de^{(s-1)}\over\longrightarrow~
  \cF_{\La}^{(s)}~\longrightarrow~ 0\,,
$$
where $s=|\De_+|$ and
\eqn\eqres{
\cF_{\La}^{(i)} ~=~ \bigoplus_{w\in W,\, \ell(w)=i}\ F_{w(\La+\rh)-\rh}\,,
}
which gives a resolution of the irreducible module $L(\La)$.
The differential of the complex is constructed from the
so-called ``screeners.''  For $\slth$
the screeners are given by
\eqn\scree{
\eqalign{
s_{\al_1} & ~=~ -\beo + \gatw \beth\,, \cr
s_{\al_2} & ~=~ -\betw\,, \cr
s_{\al_3} & ~=~ -\beth \, .\cr}
}

Applying this resolution, we may proceed with the  usual manipulations
on the ensuing double complex $(\cF\otimes F^{bc} \otimes
F^{\si\om},d,\de)$.  The first spectral
sequence associated with this double complex collapses at the second
term,
\eqn\sso{ \eqalign{
E_\infty^{p,q} ~\cong~ E_2^{p,q} & ~\cong~
  H^p(\bfnp,H^q(\delta,\cF)\otimes\mywedge\bfbm) \cr
& ~\cong~ \delta^{q,0}\, H^p(\bfnp,L(\La)\otimes \mywedge\bfbm) \,, \cr}
}
and produces the cohomology that we want to compute, while
the $E_2'$-term for the second spectral sequence is given by
\eqn\sstw{
E'_2{}^{p,q} ~\cong~ H^q(\delta,H^p(\bfnp,\cF\otimes \mywedge\bfbm))\,.
}
Let us now restrict to the case of $\slth$.
To proceed it is convenient to first make a similarity transformation
on $d$ as follows:  Introduce the operators
\eqn\eqsim{ \eqalign{
X & ~=~ -\gao\cco\bo - \gatw\cctw\btw - \gath (\cco + \cctw) \bth -
  \gao\gatw\cctw\bth \,,\cr
Y & ~=~ \gao\ctw\bth - \gatw\co\bth \,, \cr}
}
then
\eqn\ssth{
e^Y e^X\, d\, e^{-X} e^{-Y} ~=~ \si^\al e_\al - \tco\tctw\tbth \,.
}
This shows
\eqn\ssfo{
E'_1{}^{p,q} ~\cong~ H^p(\bfnp,\cF\otimes\mywedge\bfbm) ~\cong~
  H^p(\bfnp,\cF^{(q)}) \otimes\mywedge\bfbm  ~\cong~
  \de^{p,0} \, \bigoplus_{w\in W\, \ell(w)=q}
  \left( \CC_{\,w(\La+\rh) - \rh} \otimes \mywedge\bfbm\right) \, .
}
Having done this, at each Fock space in the original resolution we simply
have a copy of $\mywedge\bfbm$, and we must calculate
the cohomology of the similarity-transformed $\de$ on such.
The operators $\delta^{(i)}$, in turn, are made up of
similarity-transformed screeners.  Since the screeners now
operate on states with no $\ga$'s, we
drop their $\be$ dependence in writing the transformed result below
($s_i = e^Y e^X s_{\al_i} e^{-X} e^{-Y}$).
\eqn\simscr{
\eqalign{
s_1 & ~=~ \cco\bo - \ctw\bth\,, \cr
s_2 & ~=~ \cctw\btw + \co\bth \,,\cr
s_3 & ~=~ (\cco + \cctw)\bth \, . \cr
}
}
Explicitly, the resolution is given by
$$ \comdiag\matrix{
&& F_{r_1} & \mape{} & F_{r_1r_2}\cr
& \mapne{} &&&& \mapse{}\cr
F_\La & && \mapcr{}{} & && F_{r_1r_2r_1}\cr
& \mapse{} &&&& \mapne{}\cr
&& F_{r_2} & \mape{} & F_{r_2r_1}\cr}
$$
where $F_w \equiv F_{w(\La+\rh)-\rh}$ and the intertwiners $Q_{w,w'}\,:\,
F_w \to F_{w'}$ are given by
\eqn\eqint{ \eqalign{
Q_{w,r_i w} & ~=~ s_i^{(\La+\rh,w\al_i)}\,,\qquad {\rm if}\ \ell(r_i w)
 = \ell(w) + 1 \,,\cr
Q_{r_1,r_1r_2} & ~=~ \sum_{0\leq j\leq l_2}\ b(l_2,l_1+l_2;j) (s_2)^{l_2-j}
  (s_3)^j(s_1)^{l_2-j}\,,\cr
Q_{r_2,r_2r_1} & ~=~  \sum_{0\leq j\leq l_1}\ b(l_1,l_1+l_2;j)(s_1)^{l_1-j}
(-s_3)^j(s_2)^{l_1-j}\,, \cr}
}
where $l_i=(\La+\rh,\al_i)$ and
\eqn\eqintz{
b(m,n;j) ~=~ {m\,!\,n\,!\over j\,!\,(m-j)\,!\,(n-j)\,!}\,.
}
It is now clear that since $s_i^n = 0$ for $i=1,2,\,n\geq3$, and
$s_3^n = 0$ for $n\geq 2$ as well as $s_i^ns_3 = s_3 s_i^n = 0$
for $i=1,2,\, n\geq2$, all the differentials $\de^{(i)}$ on $E_1'$
vanish if $(\La,\al_i)\geq2$ for $i=1,2$. Thus the spectral sequence
collapses at $E_1'$, and leads to a result
consistent with that of Theorem \thNPa.
In the remaining cases the spectral sequence collapses at $E_2'$
and the result is obtained by straightforward algebra.
To formulate the answer in an elegant way let us
introduce an extension $\widetilde W$ of the Weyl group $W$ of $\slth$
by $\widetilde{W} \equiv W \cup \{ \si_1,\si_2\}$, and extend the
length function on $W$ by assigning $\ell(\si_1)=1$ and $\ell(\si_2)=2$.
Let $\widetilde{W}$ act on $\bfh^*$ by defining $\si_i\la=0,\,i=1,2$.
We can now parametrize the weights in $\mywedge \bfnm$ by $\si\in
\widetilde W$ as follows
\eqn\eqNPz{
P( \mywedge^n \bfnm ) ~=~ \{ \si \rh - \rh\,|\, \si\in \widetilde{W},
  \ell(\si) = n\} \,.
}
We now have

\thm\thNPc
\proclaim Theorem \thNPc.  For $\bfg \cong \slth$, the cohomology
$H(\bfnp,L(\La) \otimes \mywedge \bfbm)_{\La'}$ is nontrivial
only if there exists a $w\in W$ and $\si\in\widetilde W$ such
that $\La' = (w(\La+\rh)-\rh) + (\si \rh - \rh)$.  The set of
of allowed pairs $(w,\si)$ depends on $\La$ and is given in the table below.
For each allowed pair $(w,\si)$ there is a quartet of cohomology
states at ghost numbers $n, n+1, n+1$ and $n+2$ where $n=\ell(w) +
\ell(\si)$. \par \medskip\bigskip

\tbl\tballow

\begintable
\quad $w\backslash \si$ \quad | \quad $1$ \quad | \quad $r_1$ \quad |
\quad $r_2$ \quad |\quad $\si_1$ \quad |\quad $r_1r_2$ \quad |
\quad $r_2r_1$ \quad |\quad $\si_2$ \quad |\quad $r_1r_2r_1$ \quad\cr
\quad $1$   \quad    | -- | $m_1\geq1$ | $m_2\geq1$ | $m_1\geq1, m_2\geq1$ |
  $m_1\geq2$ | $m_2\geq2$ | -- | $m_1\geq1, m_2\geq1$ \nr
\quad $r_1$ \quad    |  -- | -- | $m_1\geq1$ | $m_2\geq1$ | $m_2\geq1$ |
  $m_2\geq1$ | $m_1\geq1$ | -- \nr
\quad $r_2$ \quad    | -- | $m_2\geq1$ | -- | $m_1\geq1$ | $m_1\geq1$ |
  $m_1\geq1$ | $m_2\geq1$ | -- \nr
\quad $r_1r_2$ \quad | -- | -- | $m_1\geq1$ | $m_1\geq1$ | -- |
  $m_1\geq1$  |  $m_2\geq1$ | --  \nr
\quad $r_2r_1$ \quad | -- | $m_2\geq1$ | -- | $m_2\geq1$ | $m_2\geq1$ |
  -- | $m_1\geq1$ | -- \nr
\quad $r_1r_2r_1$ \quad | -- | $m_1\geq2$ | $m_2\geq2$ | -- | $m_1\geq1$ |
  $m_2\geq1$ | $m_1\geq1, m_2\geq1$ | --
\endtable
\medskip
\centerline{Table \tballow.  Condition on $\La$ for the pair $(w,\si)$
to be allowed }
\centerline{($ m_i = (\La,\al_i)$ and -- means there's no condition
on $\La\in P_+$).}
\bigskip

For $\La$ in the bulk, the quartet structure
of $H(\bfnp,L(\La) \otimes \mywedge \bfbm)$ corresponds to the
decomposition
\eqn\eqNPw{
H(\bfnp,L(\La) \otimes \mywedge \bfbm) ~\cong~
 \left( H(\bfnp,L(\La)) \otimes \mywedge \bfnm \right) \otimes
 \mywedge \bfh\,.
}
It is quite remarkable that the quartet structure persists even away
from the bulk because, e.g., it is not true in general that
$H(\bfnp,L(\La) \otimes \mywedge \bfbm)\cong
H(\bfnp,L(\La) \otimes \mywedge \bfnm) \otimes \mywedge \bfh$.


\newsec{Comparison to $\cW$-cohomology}

Consider the $\cW$-algebra $\cW[\bfg]$ associated to the simple,
simply-laced, finite dimensional Lie algebra $\bfg$ by means
of the quantized Drinfel'd-Sokolov reduction with respect
to the principal $\sltw$ embedding in $\bfg$ (see, e.g.,
[\BS] and references therein).  We will be mainly interested
in the cases $\cW_n,\,n=2,3$, where $\cW_n \equiv \cW[\frak{sl}_n]$.
The algebras $\cW[\bfg]$ have a realization in terms of $\ell$
free scalar fields coupled to a background charge $\al_0$.  The
corresponding $\cW[\bfg]$-modules are the Fock spaces $F(\La,\al_0)$.
They are parametrized by the background charge $\al_0$ (in terms of
which the central charge is given by $c =  \ell - 12 \al_0{}^2 |\rh|^2$)
and a $\bfg$-weight $\La$. \medskip

Physical states of $\cW[\bfg]$ gravity coupled to $\ell$ free scalar
fields (i.e.\ $\cW[\bfg]$ matter) are given by the semi-infinite
cohomology of $\cW[\bfg]$ with coefficients in $F(\La^M,\al_0^M) \otimes
F(\La^L,\al_0^L)$, i.e.\ $H(\cW[\bfg], F(\La^M,\al_0^M) \otimes
F(\La^L,\al_0^L))$, where $F(\La^M,\al_0^M)$ and $F(\La^L,\al_0^L)$
represent the matter and gravity sector, respectively.
A necessary condition for the semi-infinite cohomology to
be defined is $(\al_0^M)^2 + (\al_0^L)^2 = -4$. \medskip

A particularly interesting case, commonly
refered to as the noncritical $\cW[\bfg]$ string,
is when the background charge in the matter sector vanishes
(this corresponds to $c^M = \ell$).  A subsector of this
$\cW[\bfg]$ string, defined by restricting the momenta $(\La^M,-i\La^L)$
to lie on the lattice
$L\equiv \{ (\la,\mu) \in P \times P\,|\, \la-\mu\in Q\}$, possesses
the structure of a BV-algebra.  More precisely, let $\cC$ be the
complex
\eqn\eqNPaa{
\cC \equiv \bigoplus_{ (\La^M, -i\La^L) \in L}\
  F(\La^M,\al_0^M) \otimes F(\La^L,\al_0^L) \otimes F^{gh}\,,
}
where $F^{gh}$ denotes the Fock space of the $\cW[\bfg]$ ghosts.
Note that by summing the matter momenta over the integral weight
lattice $P$, we have arranged that $\cC$ becomes a $\whg$-module at
level $1$.  In fact,
the chiral algebra $\fC$ corresponding to $\cC$ can be equipped with the
structure of a Vertex Operator Algebra.
The chiral algebra $H(\cW[\bfg],\fC)$ corresponding to the semi-infinite
$\cW[\bfg]$-cohomology of the complex $\cC$ inherits the structure of a
BV-algebra, where the product is given by the normal ordered product of
operators and the BV-operator is given by $b_0^{[2]}$ -- the zero mode
of the anti-ghost corresponding to the Virasoro subalgebra of $\cW[\bfg]$.
In addition, $H(\cW[\bfg],\fC)$ is a $\bfg^M \oplus \bfh^L$ module, where
the $\bfg$ structure is a remnant of the $\whg$-module structure
in the matter sector and $\bfh$ is the action of the Liouville momenta.
\medskip

The main result of this letter is the following

\thm\thNPg
\proclaim Conjecture \thNPg. We have an isomorphism of BV-algebras
\eqn\eqNPab{
H(\cW[\bfg],\fC) ~\cong~ H(\bfnp, \Ep(G) \otimes \mywedge \bfbm)
  ~\equiv~ \fA[\bfg] \,.
}
\par

We have checked that the conjecture is true for $\bfg\cong \sltw$, i.e.\
for the Virasoro algebra, where the cohomology on the left hand side
of \eqNPab\ was computed in [\LZvir,\BMPvir]
and the BV-structure was unraveled
in [\LZbv], by explicit comparison.  For $\bfg \cong \slth$ both the
cohomology and the BV-structure on the left hand side of \eqNPab\
were determined in [\BMPbig].  Again we find complete agreement with
Conjecture \thNPg.  Note that the fact that $\fA[\bfg]$ is cyclic
with respect to the BV-operator $\De$ (Conjecture \KPconj) is crucial
in making the identification $\De = b_0^{[2]}$.

To compare the cohomologies one has to remember
that the action of $\bfh^L$ on the left hand
side of \eqNPab\ is conventionally identified
in the literature with the action of
$-w_0(\Pi_i)$ on the right hand side.  Also, to compare the $\slth$
result with Table 3.2 in [\BMPbig] one has
to substitute $w\to w^{-1}$ and $\si \to w^{-1} \si w_0$.

\vfil\eject
\leftline{\bf Acknowledgements}

P.B.\ and J.M.\
acknowledge the support of the Australian Research Council, while
K.P.\ is supported in part by the U.S.\ Department of Energy Contract
\#DE-FG03-84ER-40168.  While finishing this paper we were informed by
G.~Zuckerman that he has obtained results in collaboration with
B.~Lian that partially overlap with ours.

\footatend\immediate\closeout\rfile
\baselineskip=14pt{\bigskip\noindent {\bf References}}%
\bigskip{\frenchspacing%
\parindent=20pt\escapechar=` \input refs.tmp\vfill\eject}\nonfrenchspacing
\vfil\eject\end